# Interaction of a Long Alkyl Chain Protic Ionic Liquid and Water

Enrico Bodo,[1] Sara Mangialardo,[2] Francesco Capitani,[2] Lorenzo Gontrani,[1,3] Francesca Leonelli,[1] Paolo Postorino[2,a]

[1]Department of Chemistry, University of Rome "La Sapienza", P. A. Moro 5, 00185 Rome, Italy

[2]Department of Physics, University of Rome "La Sapienza", P. A. Moro 5, 00185 Rome, Italy

[3]CNR-ISM, Istituto di Struttura della Materia, Via del Fosso del Cavaliere 100, 00133 Roma, Italy

A combined experimental/theoretical approach has been used to investigate the role of water in modifying the microscopic interactions characterizing the optical response of butyl-ammonium nitrate (BAN) water solutions. Raman spectra, dominated by the signal from the protic ionic liquid, were collected as a function of the water content, and the corresponding spatial organization of the ionic couples, as well as their local arrangement with water molecules, was studied exploiting *classical molecular dynamics* calculations. High quality spectroscopic data, combined with a careful analysis, revealed that water affects the vibrational spectrum BAN in solution: as the water concentration is increased, peaks assigned to stretching modes show a frequency hardening together with a shape narrowing, whereas the opposite behavior is observed for peaks assigned to bending modes. Calculation results clearly show a nanometric spatial organization of the ionic couples that is not destroyed on increasing the water content at least within an intermediate range. Our combined results show indeed that small water concentrations even increase the local order. Water molecules are located among ionic couples and are closer to the anion than the cation, as confirmed by the computation of the number of H-bonds which is greater for water-anion than for water-cation. The whole results set thus clarifies the microscopic scenario of the BAN-water interaction and underlines the main role of the extended hydrogen bond network among water molecules and nitrate anions.

---

[a] Author to whom correspondence should be addressed. Electronic mail: paolo.postorino@roma1.infn.it

## I. INTRODUCTION

Ionic liquids (ILs) are a wide class of purely ionic compounds, actually salts, characterized by a low melting temperature, typically below 100°C. As a matter of fact, anion-cation couples are made of charged molecular species whose large size remarkably attenuate the Coulomb interaction and allow for a liquid phase around or close room temperature. The dependence of the chemical-physical properties of ILs by the anion/cation size and chemical composition makes ILs interesting as a "tailorable" solvents, since the cations and anions can be chosen over a huge variety of molecular morphologies to develop a task-specific IL. Therefore, ILs are particularly suitable for many different technological and industrial applications,[1] including their use as organic solvents, lubricants, reaction and extraction media in analytical chemistry.[2,3]

In recent years, it has been established that many ionic liquids are nanostructured.[4] Cations, usually formed by an ionic and alkyl component, and inorganic anions are ordered on a length scale that is roughly given by the distance between two polar groups separated by the alkyl chain. Cations drive the segregation of the ionic liquid in polar and nonpolar domains, in a similar way to that of amphiphile self-assembly.[5]

While in micellar self-segregated liquid the order scale is nanometric, in the case of ILs we have an intermediate range order, since the length scale is around 7-15 Å. This spatial scale can be easily modified acting on the cation alkyl chains, simply increasing their length while maintaining little interdigitation, such as recently observed for the protic ionic liquid series: ethyl-, propyl-, butyl- and pentyl-ammonium nitrate.[6] Governing the intermediate range order is crucial to control the interaction of ILs with other solvents,[7] since the presence of large non polar domains greatly enhances the ability of ILs to dissolve non polar species. For this reason, and considered that many ILs are completely miscible in water,[8,9,10] it is interesting to study the effect of water as solvent. This argument is especially important if a particular subset of ILs, *i.e* protic ionic liquids (PILs) is considered. PILs are prepared through proton transfer from a Brønsted acid to a Brønsted base and are thus made by of ionic couples where proton acceptor and donor sites generate an extended hydrogen bond network that significantly affects the intermediate range order and influences a number of physical properties.[11,12]

Physical and chemical properties of generic ILs with different cation length and/or anion type with several solvents, including water, have been theoretically and experimentally investigated in the last years. In particular vibrational spectroscopies such as Raman and Infrared,[13,14,15,16,17,18] as well as structural probes[19], often in conjunction with specific theoretical simulation approaches,[20,21,22,23] have been exploited. On these basis several microscopic scenarios have been proposed although a complete and coherent picture has not yet emerged. So far, experimental observations basically converge in identifying the effects of a weak water-anion interaction for a large variety of ILs at least within a small water concentration range, in some cases so small to be considered highly diluted water. For this reason, most of the papers reporting experimental data focus on the vibrational response over the OH frequency region and thus on the water spectroscopic response[13,14,15]. DFT calculation showed that in $B_{Mim}BF_4$ and $B_{Mim}PF_6$ water solutions $H_2O$ interacts via hydrogen bonds with two anions in a symmetric structure A - - H—O—H - - A, with characteristic H-F distances of 1.88 Å and 1.98 Å for $BF_4^-$ and $PF_6^-$ respectively. However, this study is limited to low water concentration, in particular 0.5 mol/L for $B_{Mim}BF_4$ and 0.1 mol/L for $B_{Mim}PF_6$. Subsequent works, both theoretical[20] and experimental,[14] basically agree with the above depicted scenario. Experimental results[14] for other $BF_4^-$ based ILs, focused on the spectral response of the ionic couple, have evidenced a detectable blueshift of the anion F-B stretching frequency as the water-content is increased, within a concentration range of 0-5 $H_2O$ molecules per ionic couple. A spectroscopic work on IL-water solution carried out over a very large $H_2O$ concentration range [up to 56 mol/L) by Jeon *et al.*[15] suggest that the blueshift effect of the $BF_4^-$ peak is caused by structured water surrounding the anion. It is worth noting that this is also the only paper showing a similar blueshift of the cation vibrational frequencies (namely alkyl chain vibrations) on increasing the water content[15]. Finally, NMR studies and molecular dynamics simulations[22] by Moreno *et al.*

showed that there are two solvation regimes: at low water concentrations, *i.e.* below 0.25 $H_2O$ molecules per ionic couple, water interacts strongly with both cation and anion, perturbing the short-range network of the IL; at higher concentrations a nonselective interaction takes place, *i.e.* the ionic network is perturbed by the insertion of water clusters, rather than single molecules.

Despite of the amount of spectroscopic and MD data on ILs, these methods have not been systematically applied to PILs where the extent of the hydrogen bond network and, consequently, its effect on the vibrational spectrum of the ionic couple is expected to be more remarkable. In the present paper, we focus our investigation on PILs and in particular on butyl-ammonium nitrate (BAN) in water. In this case 3 hydrogen bond donors and 3 acceptors are provided by the ionic couple giving to the system the capability of building up a complex hydrogen-bonded network. Apart from this key aspect, peculiar of all PIL compounds, a further specific interest in BAN can be found in studying the perturbation induced by a growing content of water on the intermediate range order of such a long alkyl chain PIL. We thus mainly focus our spectroscopic investigation to the frequency region characteristic of the vibrational modes of the ionic couple and, exploiting the low water cross section, we carried out a systematic Raman study over a wide water concentration range up to more than 25 $H_2O$ molecules for ionic couple. Careful *molecular dynamics* calculations help us in the interpretation of experimental data and provided a deep structural analysis of the microscopic local arrangement and of the hydrogen bond network.

## II. EXPERIMENTAL AND THEORETICAL METHODS

*Materials* - To a solution of butylamine (10 mL, 0.10 mol) in pentane (10 mL), cooled at -20°C, $HNO_3$ (65% wt/wt, 7.1 mL, 0.10 mol) was cautiously added dropwise while stirring. The reaction mixture was stirred for 1h at the same temperature and after that time the pentane layer was separated, the product washed with pentane and dried at the rotary evaporator. The residual water was removed, while stirring, on standing under high vacuum pump for 72h and the water final quantity was evaluated by $^1$H-NMR analysis (H$_2$O < 0.003wt.%). The compound was kept under a nitrogen atmosphere. Data of butyl-ammonium nitrate (BAN)[24]: white solid; d(g·cm$^{-3}$) =1.07; $^1$H-NMR (DMSO-d$_6$): 7.81 (*bs*, 3H), 2.78 (*t*, *J* = 7.3, 2H), 1.57-1.42 (*m*, 2H), 1.29 (*sex*, *J* = 7.3, 2H), 0.85 (*t*, *J* = 7.3, 3H) ; $^{13}$C-NMR (DMSO-d$_6$): 38.9, 29.3, 19.3, 13.6. All the BAN/H$_2$O solutions were prepared under a nitrogen atmosphere and the solution resulting densities were volumetrically determined (error of ±0.02 g·cm$^{-3}$).

*Raman measurements* - Raman measurements were carried out using a confocal-microscope Labram Infinity spectrometer by Jobin Yvon, equipped with a set of interchangeable objectives with long working distances and different magnifications from 4× to 100×. The samples were excited by the 632.8 nm line of a 30 mW He-Ne laser. The 1800 lines/mm grating monochromator and a charge-coupled-device (CCD) detector allowed for a spectral resolution better than 3 cm$^{-1}$. Elastically scattered light was removed by a holographic notch filter, which avoided collection of spectra within the low-frequency spectral region. Raman spectra were indeed collected in backscattering geometry over different spectral ranges spanning from 300 to 4000 cm$^{-1}$. The Raman spectra were collected in thin quartz cuvettes, the use of a rather small confocal diaphragm (<100 µm) combined with a 50x objective allowed us to minimize the contribution arising from the quartz windows. To cover all the spectral ranges, for each samples 4 spectra were acquired. Indeed, the spectra in the low frequency region presented in this paper were obtained combining two different spectral windows. Further experimental details on the apparatus can be found in Ref. 25.

From a preliminary set of measurements we observed that most of the shifts in the Raman spectra were in the range of a few cm$^{-1}$ on going from the neat BAN to the most diluted solution. In order to obtain the maximum precision in comparing spectra, avoiding any possible effect of frequency calibration, we decided to repeat the measurements acquiring at a fixed spectral window the spectra from all the solution investigated before changing the spectral range, keeping constant the other experimental parameters. This procedure has made us confident even about the smallest frequency shifts (~ 1 cm$^{-1}$) observed.

*Molecular dynamics simulation* - Molecular dynamics was performed on 3 different BAN/water mixtures with molar ratios 1:2, 1:9 and 1:20 plus a simulation on neat BAN. In each simulation the number of BAN ionic couples was set to 100 and the liquid cell was complemented using the correct number of water molecules. The cells were created by placing randomly the molecular constituents and then compressed without electrostatic interaction so to allow a complete mixing of the constituents. The resulting samples, now with electrostatic turned on, were then equilibrated in a NPT *ensemble* at room temperature and pressure for 2 ns up to a constant density. Production and analysis was performed for 10 ns on each sample using an NVT ensemble. No bonds were constrained and the integration time step was 1 fs. A total of 10 ns was used for production. All the calculations have been performed using DL_POLY [26] while part of the analysis have been performed using TRAVIS[27] and "in house" codes.

The force field used for BAN is a trivial complementation of the one used by us in Ref. 28 on EAN that is, in turn, a derivation of the the OPLS force field [29]. The force field for water is the TIP4P. A comparison of the perfomances of the employed force field with ab-initio molecular dynamics has been recently published by us.[30]

The adopted force field has been able to reproduce the experimental, measured density of the samples accurately enough as shown in Table I.

Table I. Theoretical and measured density of a selected set of the mixtures.

| Molar ratio Water/BAN (theory) | Density (theory) gr/cc | Molar ratio Water/BAN (mixture) | Density (measured) gr/cc (Ban molar fraction) |
|---|---|---|---|
| 0 | 1.106 | 0 | 1.105 |
| 2 | 1.103 | 2.03 | 1.092 |
| 9 | 1.06 | 7.40  | 1.066 |
| 20| 1.05 | 21.36 | 1.029 |

A scheme of the atom types of BAN along with the naming conventions in the OPLS force field is reported in Figure 1.

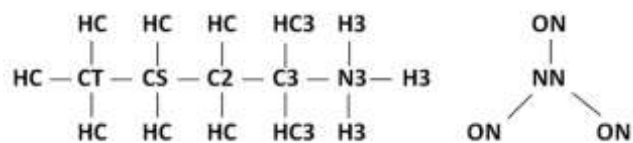

FIG. 1. Atom type used in OPLS atomic force field calculations. Atomic labels of the butil-ammonium cation are shown on the left and those of nitrate anion on the right.

## III. RESULTS AND DISCUSSION

### A. Raman results

Since water is a weak Raman scatterer, the spectra collected from the different solutions are expected to be basically dominated by the Raman signal from BAN. For the sake of simplicity we use a parameter R for the water content given by the number of water molecules ($n_{H_2O}$) over the number of BAN ionic couples ($n_{BAN}$), $R = n_{H_2O}/n_{BAN}$. The Raman spectra collected from 12 samples with R varying over the 0-26 range are shown over the low and the high frequency regions in panel (a) and (b) of Figure 2 respectively. Following Ref. 31, the assignments of the most relevant Raman peaks are also reported.

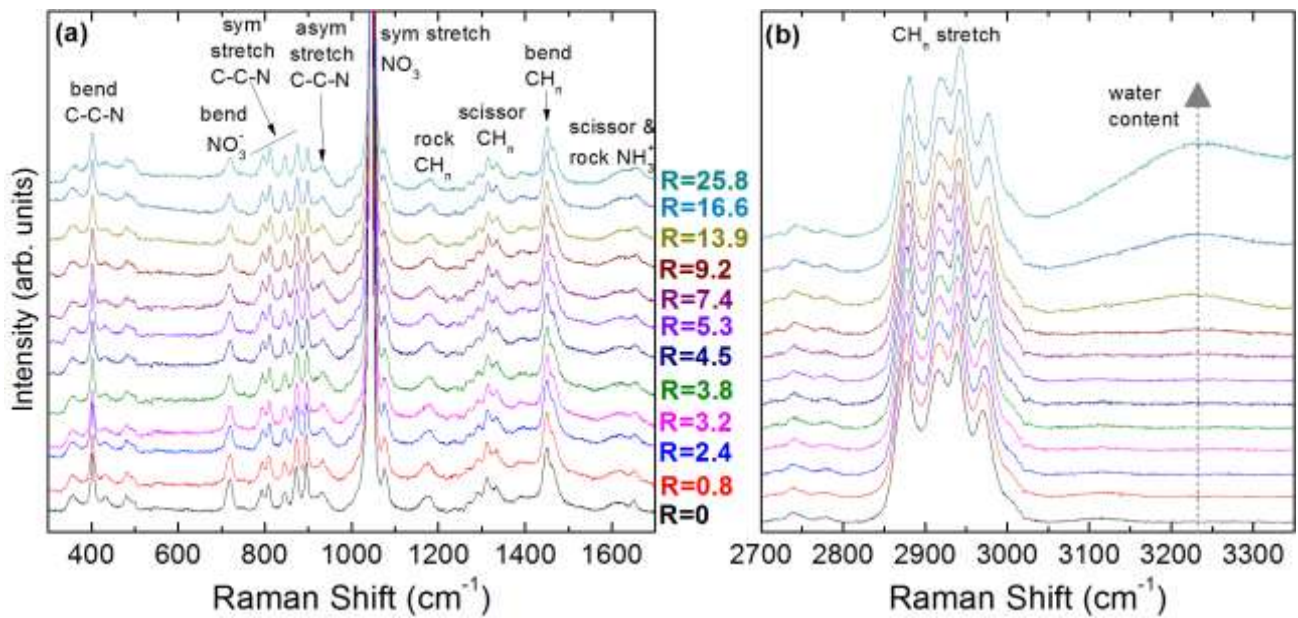

FIG. 2. Raman spectra of BAN at different water content (R). The number of water molecules respect to BAN ionic couples increases from down to top. (a) Low frequency region, (b) CH region. The arrow in (b) highlights the increasing water content. Peak labelling is from Ref. 31

As expected, the spectra are dominated by the signal from BAN, albeit looking in Figure 2(b) the increasing content of water can be monitored looking at the peak centered around 3240 cm$^{-1}$ assigned to vibrational modes of liquid water.[32]

Data reported in Figure 2 show that water modifies the Raman spectra of BAN in a very weak and continuous way. The only apparent change in the spectra is in the frequency range around 1600 cm$^{-1}$, but this is due to the superimposition of the scissoring and rocking vibration of $NH_3^+$ on the bending vibration of water,[33] whose intensity is increasing with the water content. All the peaks labeled in Figure 2 were carefully analyzed following a standard fitting procedure previously applied to Raman spectra.[34] In particular, in the present case, Lorentzian peak shapes have been used. The fitting procedure thus provided three fitting parameter values for each analyzed peak, *i.e.* the central peak frequency $\nu$, the linewidth and the intensity. Among all the peaks analyzed, only those reported in the ten panels of Figure 3 show a clearly detectable water content dependence of the central frequencies.

In particular, for these peaks, a more or less steep hardening of the central frequency has been observed on increasing of the water content whatever the peak originates from the cation or the anion moieties. Looking at the R-dependence of the peak frequencies shown in Figure 3 we tried to reproduce the experimental behavior with an exponential decay in the form:

$$\nu(R) = \nu_0 + \Delta\nu * [1 - e^{-\frac{R}{\rho}}]$$

where $\Delta\nu$, the frequency variation, and $\rho$, the exponential decay constant, are the free fitting parameters and $\nu_0$ is the peak frequency measured for the pure BAN ($R = 0$). Applying a standard fitting procedure we obtain good agreement with a single exponential decay for all the peaks but the anion stretching mode for which only the addiction of a second exponential decay allows to reproduce with a good accuracy the experimental data. In this way, a very good agreement between fitting functions and experimental data has been found for all the analyzed peaks, as shown in Figure 3, where experimental data and best fit curves are compared. The best fit parameter values are reported Table II. It is interesting to notice that the frequency variation is actually the same for all the vibrational modes whereas the additional decay constant ρ* obtained for the anion peak relative to the NO$_3$ symmetrical stretching mode (*red* in Figure 3 and Table II) is much smaller than all the others representing a *fast* anion-water interaction process i.e. induced by a rather small amount of water. It is worth noting that a previous molecular dynamics study shows that mixtures of a non-protic ionic liquid with water also have two distinct regimes of solvation[22]. Since our analysis show that only the anion stretching vibration shows a double exponential behavior, our results indicate that this mode is affected by the interaction with water molecules in almost the same way and can be seen representative of the building up of the hydrogen bond network. As we will see below this effect is in line with what we can conclude from the structure of the mixtures as determined theoretically where the anion is coordinated by water via H-bonds much more than the cation.

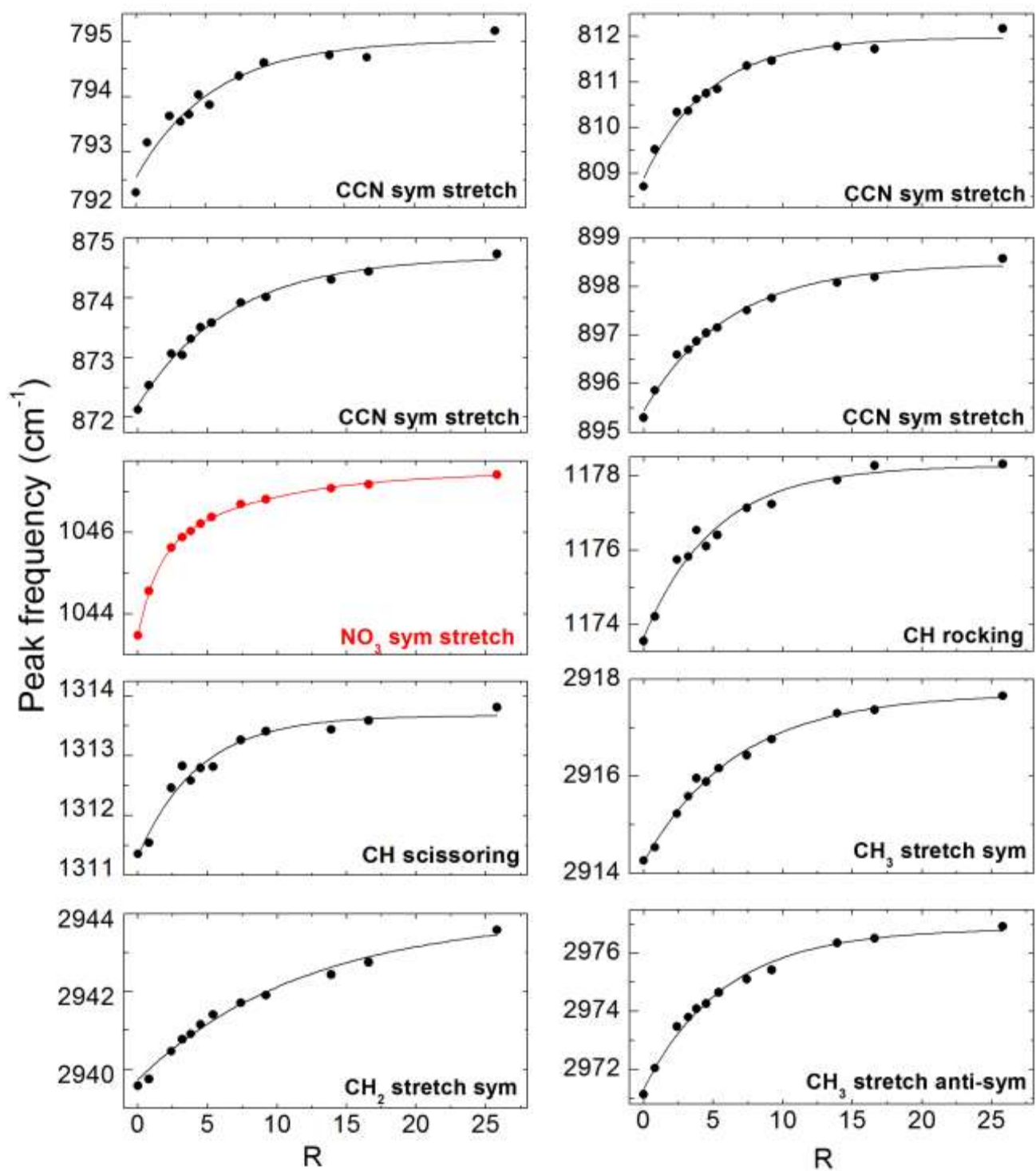

FIG. 3. Frequency behavior of the peaks that exhibit detectable R-dependence. The line in each panel is the best fit. The panel in red highlights the only peak ascribed to an anion vibrational mode.

TABLE II. Best fit parameter values obtained using the exponential fitting curve reported in the text. A double exponential fitting curve is appropriate for the anion vibrational mode only, the relative best fit parameters are reported in red and values marked with the asterisk refer to the second exponential curve.

| $\nu_0$ / Assignment | $\Delta\nu$ (cm$^{-1}$) | $\rho$ |
|---|---|---|
| 792 cm$^{-1}$ / sym stretch CCN | 2,5(2) | 5.6 (10) |
| 809 cm$^{-1}$ / sym stretch CCN | 3,1 (1) | 4.7 (5) |
| 872 cm$^{-1}$ / sym stretch CCN | 2,5 (1) | 6.6 (6) |
| 895 cm$^{-1}$ / sym stretch CCN | 3,0 (1) | 6.1 (5) |
| <span style="color:red">1043 cm$^{-1}$ / sym stretch NO$_3$</span> | <span style="color:red">2,2 (1)</span><br><span style="color:red">$\Delta\nu^*$=1,8 (1)</span> | <span style="color:red">7.2 (4)</span><br><span style="color:red">$\rho^*$=1.2 (1)</span> |
| 1174 cm$^{-1}$ / rock CH$_n$ | 3,0 (2) | 5.1 (5) |
| 1311 cm$^{-1}$ / CH scissoring | 2,3 (1) | 4.3 (6) |
| 2914 cm$^{-1}$ / sym stretch CH$_2$ | 3,4 (1) | 6.7 (5) |
| 2940 cm$^{-1}$ / asym stretch CH$_2$ | 4,2 (1) | 12.0 (1) |
| 2971 cm$^{-1}$ / asym stretch CH$_3$ | 5,5 (2) | 5.8 (4) |

As mentioned above, the peaks analyzed so far are the only ones for which a clear R-dependence of the central frequencies has been found. Nevertheless a careful analysis of the effect of water on the peaks ascribed to bending modes of both anion, at about 720 cm$^{-1}$, and cation, in the range 350-500 cm$^{-1}$, allows for further relevant considerations. The central frequency and width of the peaks assigned to these bending modes are shown in Figure 4(a) as a function of R. Peak frequencies do not show a clearly definite R-dependence and, within present experimental uncertainties, they can be assumed constant overall the explored R-range. Actually, the two cation C-C-N bending modes at the lowest frequencies might also show a slightly decrease but, bearing in mind that frequencies vary by less than 1 cm$^{-1}$, it is hard to be confident of this dependence. A more clear effect can be found for the linewidth R-dependencies shown in Figure 4(b). Here a two steps behavior, increasing at low water content (0<R<7) and basically constant or weakly decreasing on further increasing the solvent, has been found.

It is worth to make a comparison among the linewidth R-dependences of the anion $NO_3$ and cation C-C-N bending modes in Figure 4(b) with those relative to the stretching modes of the same bonds, shown in Figure 5.

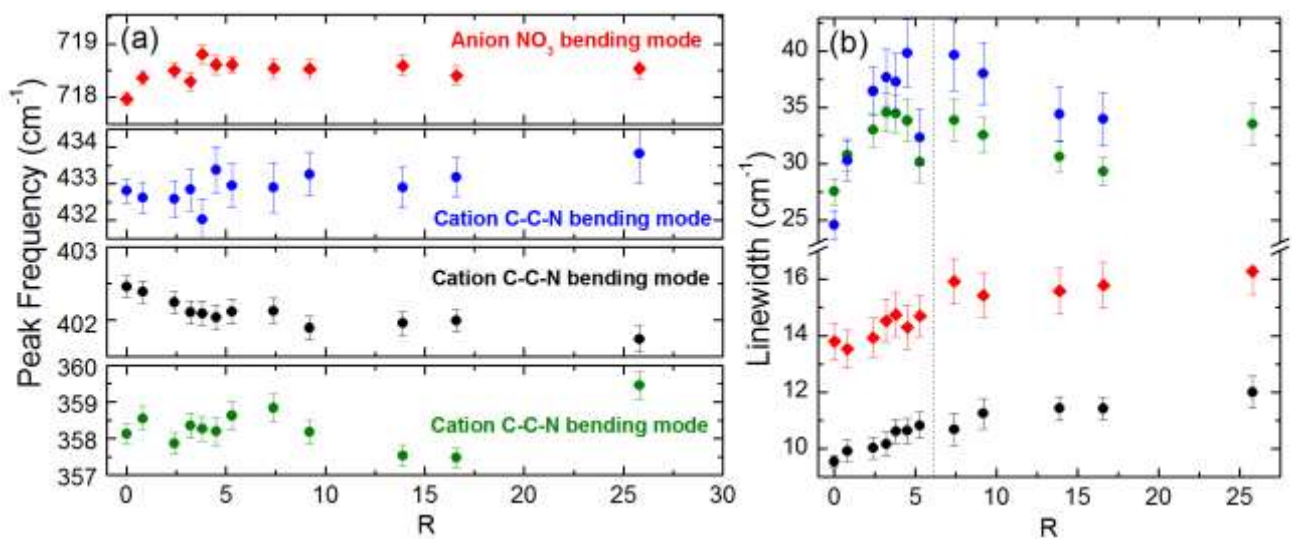

FIG. 4. Peak frequency (panel (a)) and widths (panel (b)) of both anion and cation bending modes as a function of water content. The dashed line in panel (b) marks the concentration value of the discontinuity in the linewidth behavior. The same color code has been used in panels (a) and (b).

The linewidth of the anion stretching mode shows a decreasing behavior which can be well described by a simple exponential decay (see Figure 5) with a decay constant $\rho_t \approx 6$. This concentration value compares well with the one marking the discontinuity of the linewidth R-dependence found for the bending modes (see Fig. 4(b)). On the other hand, the linewidths of the cation C-C-N stretching modes keep constant (792 cm$^{-1}$) or are weakly decreasing (809 cm$^{-1}$) as the water content is increased. Bending and stretching modes thus show different, almost opposite, R-dependence: on increasing R the former show constant or slightly decreasing central frequencies and increasing linewidths (see Figure 4) whereas the latter combine a remarkable frequency hardening with, at least for the anion stretching mode, a strong peak narrowing (see Figure 3 and 5).

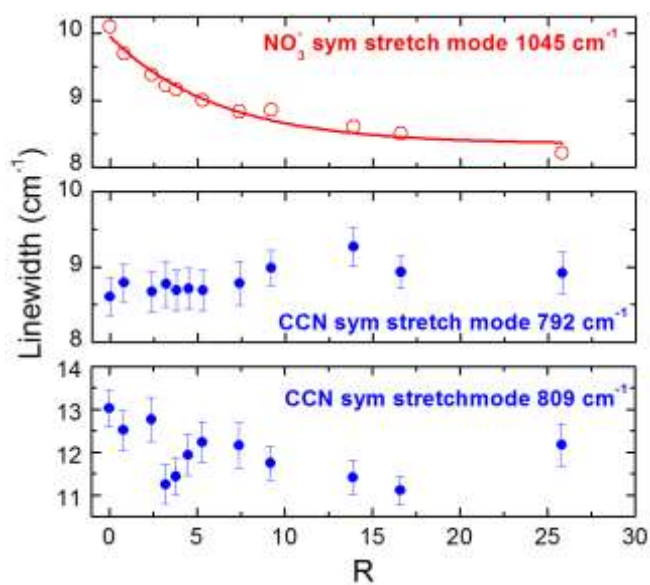

FIG. 5. Peak widths of both anion and cation symmetrical stretching vibration as a function of water concentration. In the upper panel the red line is the curve of best fit, an exponential decay function, and the error bars are within the symbol size. The corresponding peak frequency behavior has been already presented in Figure 3.

This peculiar behavior can be ascribed to the non-trivial effects related to the delicate balance among the water-anion, water-cation and ion-ion interactions. Given the nature of the solutions investigated it is expected that a key role is played by the H-bond and by the onset of extended, more or less structured, network. In particular as the number of water molecules per ionic couple, R, is increased, part of the strong anion-cation H-bonds are replaced by much weaker water-anion H-bonds. As a consequence of the progressive substitution of the strong H-bond with weaker ones the N-O bond is strengthened, ultimately explaining the blueshift of the N-O stretching line (see Figure 3). The reduction of the cation-anion interaction strength also leads to the remarkable peak narrowing (see Figure 5, upper panel) observed for the anion $NO_3$ stretching.

On the other hand, for the anion $NO_3$ bending mode the introduction of a weaker H-bond affect mainly the linewidth, rather than the vibrational frequency (see Figure 4). The different behavior shown by the bending and stretching modes on increasing the water content can be ascribed to the strong directionality of the H-bond. Indeed, for the anion stretching modes the oxygen vibration basically occurs along H-bond whereas for bending modes the vibration is mainly transversal to the H-bond direction. For this reason a decrease in H-bond strength tend to manifest with a redshift of the corresponding bending frequencies.

**B. Computational results**

To better understand the relative arrangement of BAN and water, four molecular dynamics simulations were done to replicate the BAN/water mixtures behavior for R equal to 0, 2, 9 and 20. In Figure 6 representative snapshots of the molecular dynamic simulations are shown highlighting the different atom types in order to provide a pictorial view of the inherent structure of the mixture. The neat liquid is shown at the top and the lower panels report the snapshot for gradually more diluted mixtures. The first column shows the carbon and nitrogen atoms of the cation, the second all the nitrogen atoms and the third the nitrogen atoms plus the water oxygen. The panels showing the carbon atoms of the cations provide a clear pictorial view of the local fluctuations of their density in the cell that can be ascribed to a segregation effect between the alkyl chains which form aggregated structures that are responsible for the intermediate range order. This order is preserved in the presence of a small amount of water (R=2) and possibly, even increased as we will see below. High concentrations of water, on the other hand (R=9 and 20) tend to partially destroy the regular patterns seen in the less concentrated mixtures, without however being able to totally destroy the cooperation between alkyl chains.

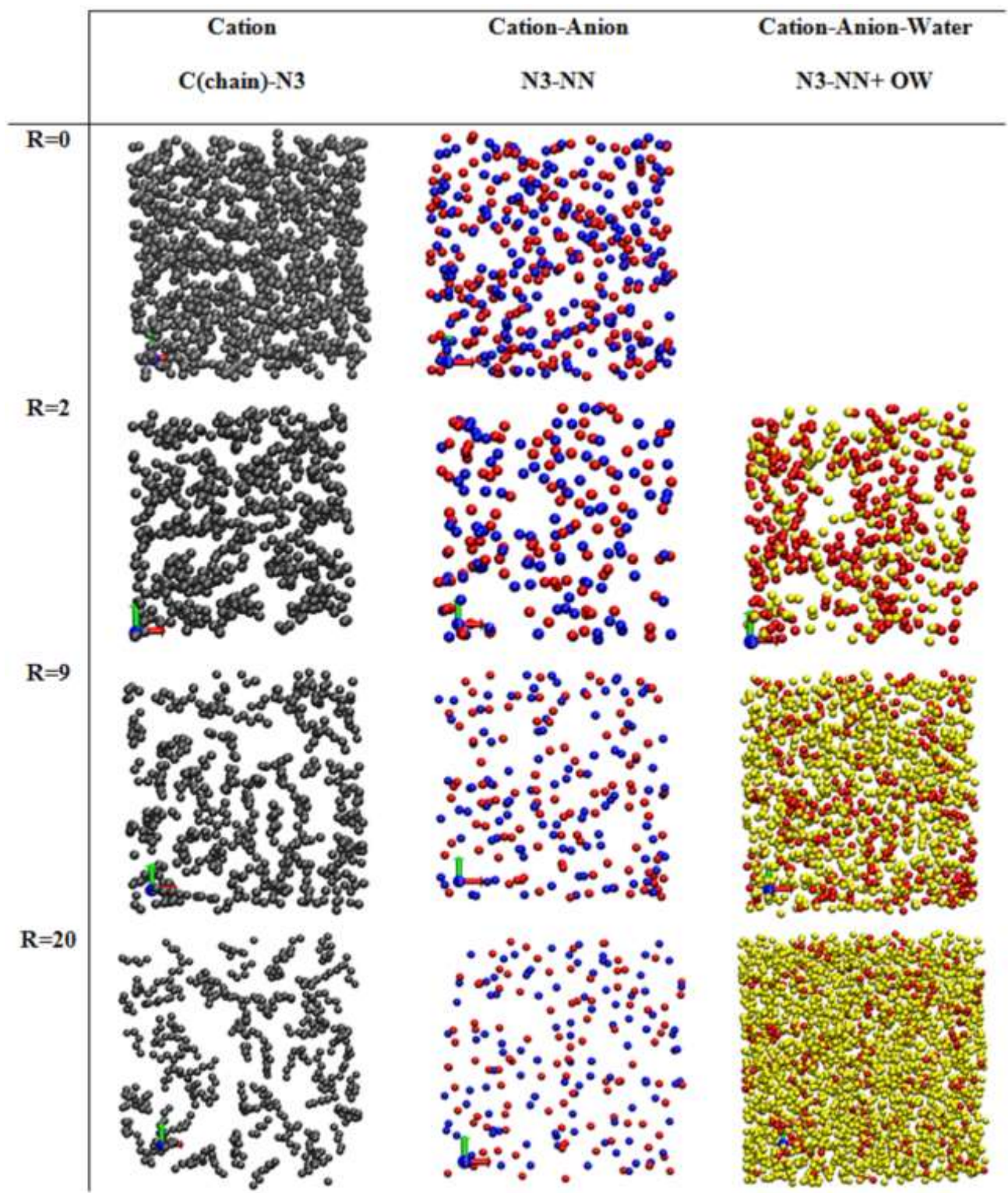

FIG. 6. Snapshots of the simulations. From top to down we find snapshots from neat BAN to R=20 solution, while from left to right the snapshots of the entire cation (gray), anion + cation (NN in red and N3 in blue), anion + water (NN in red, OW in yellow) are shown. See Figure 1 for the definition of each atom label.

If we now look at the relative position of the nitrogen atoms of the cation and anions (second column in Figure 6) we see that most, if not all, the ionic couples survive to water mixing, even at the R=9 dilution investigated. Few isolated anions and cations are found with R=20 where some of the ionic couples are fully dissociated although the degree of dissociation seems to remain relatively low. Finally, from the third column of Figure 6, where we superimpose the water oxygen atom to the anions' nitrogen atoms, we can see that water molecules tend to be segregated between the alkyl chains and to solvate the anions prominently.

From the classical trajectory it is possible to extract *S(q),* the "total (static) structure factor" that constitutes the structurally sensitive part of the X-ray scattering intensity.

The variable *q* is the magnitude of the transferred momentum, and depends on the scattering angle (2$\theta$), according to the relation $q = 4\pi \sin\theta /\lambda$. The function *S(q)* is related to the pair correlation functions descriptive of the structure, according to the formula

$$S(q) = \sum_{ij}^{N} x_i x_j f_i f_j H_{ij}(Q)$$

where we have introduced the partial structure functions $H_{ij}$ defined in terms of pair correlation functions by the Fourier integral[35]

$$H\_ij\,(Q) = 4\pi\rho_0 \int_0^\infty dr\, r^2 [g_{ij}(r) - 1] \frac{\sin qr}{qr}$$

($\rho_0$ is the bulk number density of the system, $x_i$ are the numerical concentrations of the species and $f_i$ their *q*-dependent X-ray scattering factors).

In Figure 7 the total structure factor, *S(q)*, for the four studied dilution is reported. All the peaks shown in Figure 7 for neat BAN are compatible with previous literature[17] and with in-house X-ray data (not shown here). The first peak, around 0.5 Å, has been attributed to the intermediate range order.[36] Our data are consistent with this attribution since the value obtained for the intermediate range order is exactly the distance between two polar groups, separated by the alkyl chain domain. It is possible to estimate the cation length from the Tanford equation:

$$l_{max} = 1.5 + 1.265 \cdot n_c$$

where $n_c$ is the number of carbon atoms that are in the portion of the alkyl chain that is incorporated in the hydrophobic core.[37] For BAN ($n_c = 4$) the estimated upper value is thus $l_{max} = 6.56\ \dot{A}$. The cation length from the nitrogen atom to the end of the alkyl chain can be accurately determined by the molecular dynamic output (Figure 7, inset) and it is about 5 $\dot{A}$. The peak in the *S(q)* is precisely twice these values and equal to 12 Å.

The evolution of peak1 with the growing content of water, provides information about the effect of water addition on the intermediate range structural organization of BAN and represent a more quantitative assessment of the previous discussion based on the visual analysis of the dynamics snapshots. We begin by noticing that adding small quantities of water leads to an increase in the amount of ordering, as the growth of peak1 for R=2 (red curve) suggests. On the contrary, adding a large amount of water to BAN causes a decrease of the intermediate range structuring, as shown by the intensity reduction of peak1 for R=9 (blue curve). When BAN is even more diluted in water (R=20) the ionic couples, while still associated for the most, are separated one from each other by water and therefore the segregation forces due to polar domains ceases to exists. Therefore when we reach the R=20 concentration the alkyl chain segregation effect essentially disappears.

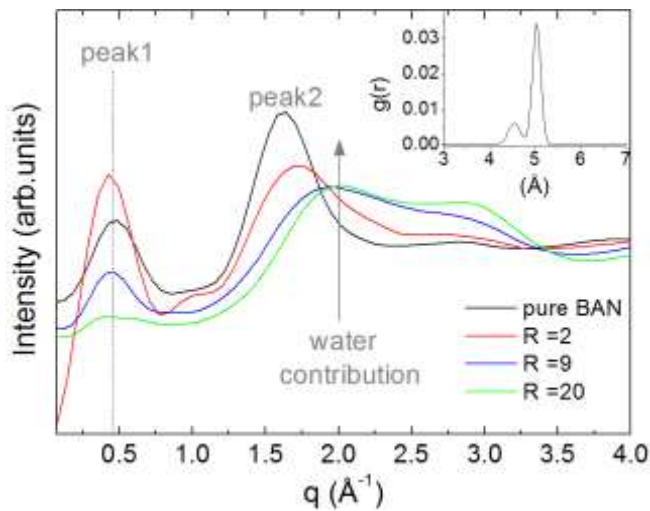

FIG. 7. Total structure factor, S(q), for the four compounds reproduced by *ab initio* calculation. Black line: neat BAN; red line: R=2, blue line: R=9 and green line: R=20. In the inset: cation length from N3 to CT (see Figure 1 for the definition of these atomic labels) estimated by molecular dynamics for neat BAN.

Looking at peak2 of the *S(q)* in Figure 7, we can see that this peak undergoes significant changes on increasing the water content. These changes are essentially due to the increasing content of water, that for a renormalization effect, takes over the stable cation-anion contribution associated to peak2.[38]

The radial distribution functions *g(r)* of the anion and cation center of mass, shown in Figure 8, confirms the stability of the ionic couple on increasing the water concentration. As a matter of fact, from the neat BAN to the most diluted solution the first shell is always clearly distinguishable, whereas the second shell completely vanishes. A clear picture arises: water places itself mostly among the ionic couples and seems to be unable to dissociate them. The disappearance of a alternating charges second shell coordination is perfectly in line with the disappearance of the peak1 in the *S(q)*. To further characterize the mixture structure the *g(r)* between the water hydrogen, the anion oxygen and the nitrate one are shown in Figure 9. In the lower panel the *g(r)* between the anion oxygen and the and cation hydrogen is also reported.

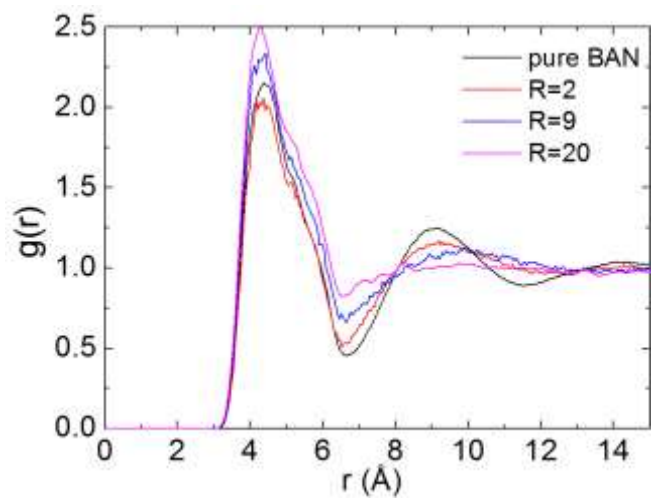

FIG. 8. Radial distribution functions of the anion-cation center of mass. See Figure 1 for the definition of each atom label.

The distance between water and ions gives us an estimate of the H-bonds that rise between the oxygen of the anion and the hydrogen of the water (ON and HW, respectively) and between the hydrogen of the cation and the oxygen of the water (H3 and OW, respectively). The *external* (outside the ionic couple) H-bond is stronger among water molecules and anions (Figure 9, a) than among water molecules and cations (Figure 9, b). Finally, looking at the *internal* (inside the ionic couple) H-bonds reported in Figure 9 (c), it is apparent that water has a minor effect on these bonds since, even at the highest dilution, the g(r) function substantially keeps its original aspect.

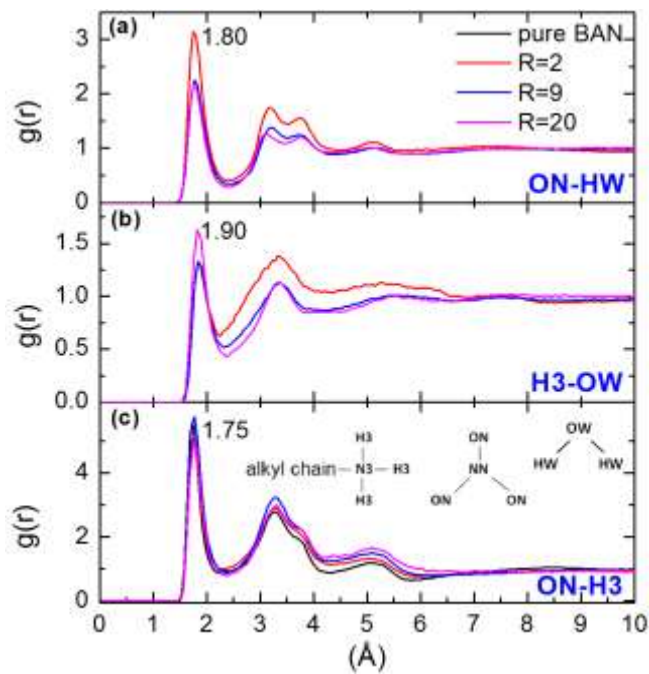

FIG. 9. Radial distribution functions for: (a) the anion-water distance (ON-HW); (b) the cation-water distance (H3-OW) and (c) the anion-cation distance (ON-H3). In panel (c) each atom label is reported as in Figure 1. In each panel, the number next to the first peak indicates the peak position.

The number of all the possible H-bonds in the water-BAN solution is shown in Fig.10 as a function of R. As mentioned above, in our simulations the number of water molecules is different at different molar ratios whereas the number of ionic couples is kept constant to 100. We had therefore to normalize the total number of H-bonds to the number of IL couples and water molecules. The number of H-bonds in the IL has been divided by 100 while the number of H-bonds with and within the water molecules has been normalized to the number of water molecules. It can be noticed that, despite the fact that ionic couples still survive even at the greatest R values, the number of anion-cation H-bonds decreases significantly going from neat BAN to R=20. In neat BAN there is an average of about two H-bonds for each anion, which abruptly decreases down to one at the lowest dilution considered (R=2). This behavior can be put into relation to the *fast* decay constant $\rho^*=1.2$ obtained for the anion symmetric stretching mode which is apparently the direct marker of this structural process. As the water content is increased, further reduction of the number of anion-cation

H-bonds is observed. The molar concentration of the IL molecules that are still connected by an H-bond between them decreases from 8.8 mol L$^{-1}$ in the neat IL to 0.53 8 mol L$^{-1}$ at the highest dilution (R=20). At the same time, the number of water-water H-bonds increases and the system naturally evolves to neat water,[39] as also shown by the *g(r)* in the inset of Figure 10. It is also interesting to notice that water forms H-bonds preferentially with anions rather than cations, as the comparison between red and blue curves reveals. The molar concentration of the water-bonded anions and cations does not vary much and remains around 6 and 1 mol L$^{-1}$, respectively. Given that the dilution of the anion and cations is increasing, we have that the number of H-bonds between them and water, per water molecule, decreases slightly with increasing water content as shown in Figure 10. The structural scenario emerging from the present computational approach does agree with the experimental finding, previously discussed, of the strong sensitivity to the water content of the anion NO$_3$ symmetrical stretching mode.

Finally, it is interesting to note that the decay constant $\rho_t \cong 6$, previously identified after a thorough line-shape analysis of the anion stretching mode, is roughly coincident with the value for which the number of water-water H-bonds gets larger than that of the water-anion H-bonds (see the intersection of the green and red lines in Figure 10). Such a concentration can be seen as a threshold above which further dilution has almost negligible effects on dynamics and interactions of BAN molecules.

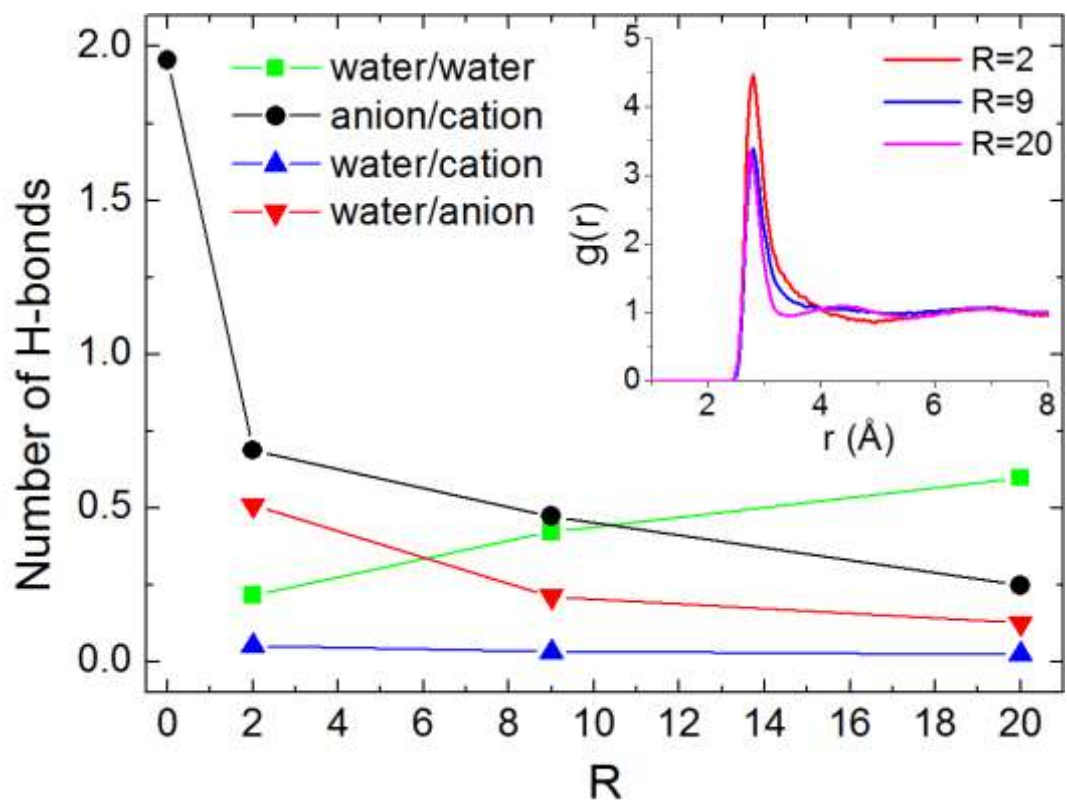

FIG. 10. Number of H-bonds (per molecule) among the ions in BAN (black), among the water molecules (green), among water molecules and anions (red) and among water molecules and cations (blue) as a function of concentration (see text for details). Inset: radial distribution functions on growing the water content for the distance of water molecules (OW-OW). See Figure 1 for the definition of each atom label.

## IV. CONCLUSIONS

BAN water solutions have been investigated by means of Raman spectroscopy and molecular dynamics simulation, as a function of water concentration. Spectroscopic data show that the Raman spectrum is dominated by the BAN signal which is weakly affect by the water content and only a thorough line-shape analysis has been able to find out dilution effects. The results obtained show that the strongest effect of water-BAN interaction can be found looking at the anion $NO_3$ symmetrical stretching mode. This peak, increasing the water content, shows a well detectable frequency hardening together with a peak narrowing. These findings have been ascribed to the progressive replacement of strong anion-cation H-bonds by weaker water-anion ones as the water content is increased. Similar effects, although with a lesser extent, have been found also for the other stretching modes analyzed whereas an almost opposite R-dependence has been found for several bending modes. This different behavior has been ascribed to the strong directionality of the H-bond which is parallel to the atomic vibration for the stretching modes and transverse for the bending modes. The latter conjectures have been basically confirmed by the structural analysis carried out following the computational approach.

Molecular dynamics simulations have provided us information about changes in the arrangement and order of BAN in water through the calculation of the total structure factor $S(q)$ and radial distribution functions $g(r)$. Simulations indicate that neat BAN is nanostructured with an intermediate range order around 12 Å and that a small amount of water (up to 2 in molar ratio) does not initially destroy such order but rather seems to further stabilize the hydrophobic segregated structures originating it. We have demonstrated that water molecules are mostly H-bonded to anions and that the number of H-bonds between the cations and the anions decreases on increasing the water concentration. We have concluded that the water presence instead of actually dissociating the ionic couples, it gently mixes with the BAN moiety and simply segregates the ionic couples one from each other. The destruction of a second shell contacts between BAN ionic couples detected both from the *g(r)* and from the shape of the *S(q)* is consistent with the progressive solvation mechanisms here explained. It is worth noting that the evidences from calculations are consistent with the observation of a particularly marked effect of water on the frequency of the anion $NO_3$ symmetrical stretching mode. In particular we have observed that characteristic decay constants appear to be related to structural rearrangements. Therefore, on one side this agreement make us confident about the theoretical results we have obtained, on the other side this means that Raman spectroscopy provides a useful and easy tool to probe the interaction between ionic liquids and water.

[1] J. Holbrey, K.R. Seddon, *Clean Products and Processes* **1**, 223 (1999).

[2] T. Welton, *Chem. Rev.* **99**, 2071 (1999).

[3] D. Constantinescu, H. Weingartner and C. Herrmann, *Angew. Chem. Int. Ed.* **46**, 8887 (2007); T. L. Greaves and C. J. Drummond, *Chem. Rev.* **108**, 206 (2008);

[4] T. L. Greaves, D. F. Kennedy, A. Weerawardena, N. M. K. Tse, Nigel Kirby and C. J. Drummond., *J. Phys. Chem. B* **115**, 2055 (2011).

[5] R. Hayes, S. Imberti, G.G. Warr and R. Atkin, *Phys. Chem. Chem. Phys.* **13**, 3237 (2011).

[6] D. F. Kennedy, S. T. Mudie and C. J. Drummond, *J. Phys. Chem. B* **114**, 10022 (2010).

[7] T. L. Greaves, D. F. Kennedy, N. Kirby and C. J. Drummond, *Phys. Chem. Chem. Phys.* **13**, 13501 (2011).

[8] S. G. Kazarian, B. J. Briscoe and T. Welton, *Chem. Commun. (Cambridge)* **2000**, *2047*.

[9] S. N. V. K. Aki, J. Brennecke and A. Samanta, *Chem. Commun. (Cambridge)* **2001**, 413.

[10] L. Cammarata, S. G. Kazarian, P. A. Salter and T. Welton, *Phys. Chem. Chem. Phys.* **3**, 5192 (2001).

[11] M. S. Miran, H. Kinoshita, T. Yasuda, M. A. B. H. Susan and M. Watanabe, *Chem. Commun.* **47**, 12676 (2011).

[12] T.L. Greaves, A. Weerawardena, C. Fong, I. Krodkiewska and C. J. Drummond, *J. Phys. Chem. B* **110**, 22479 (2006).

[13] Y. Danten, M. I. Cabaco and M. Besnard, *J. Phys. Chem. A* **113**, 2873 (2009).

[14] T. Masaki, K. Nishikawa and H. Shirota, *J. Phys. Chem. B* **114**, 6323 (2010).

[15] Y. Jeon, J. Sung, D. Kim, C. Seo, H. Cheong, Y. Ouchi, R. Ozawa and H. Hamaguchi, *J. Phys. Chem. B* **112**, 923 (2008).

[16] B. Wu, Y. Liu, Y. Zhang and H. Wang, *Chem. Eur. J.* **15**, 6889 (2009).

[17] H. Abe, Y. Yoshimura, Y. Imai, T. Goto and H. Matsumoto, *J. Mol. Liq.* **150**, 16 (2009).

[18] B. Fazio, A. Triolo and G. Di Marco, *J. Raman Spectrosc.* **39**, 233 (2008).

[19] R. Hayes, S. Imberti, G. G. Warr and R. Atkin, *Angew. Chem. Int. Ed.* **52**, 4623 (2013).

[20] C. Roth, A. Appelhagen, N. Jobst and R. Ludwig, *Chem. Phys. Chem.* **13**, 1708 (2012).

[21] A. R. Porter, S. Y. Liem and P. L. A. Popelier, *Phys. Chem. Chem. Phys.* **10**, 4240 (2008).

[22] M. Moreno, F. Castiglione, A. Mele, C. Pasqui and G. Raos, *J. Phys. Chem.* **112**, 7826 (2008).

[23] B. L. Bhargava and M. L. Klein, *J. Phys. Chem. A* **113**, 1898 (2009).

[24] D. F. Kennedy and C. J. Drummond, *J. Phys. Chem. B* **113**, 5690 (2009).

[25] S. Mangialardo, F. Piccirilli, A. Perucchi, P. Dore and P. Postorino, *J. Raman Spectrosc.* **43**, 692 (2012).

[26] W. Smith, C. Yong and P. Rodger, *Mol. Simulat.* **28**, 385 (2002).

[27] M. Kirchner and B. Brehm, *J. Chem. Inf. Model* **51**, 2007 (2011).

[28] L. Gontrani, E. Bodo, A. Triolo, F. Leonelli, P. D'Angelo, V. Migliorati and R. Caminiti, *J. Phys. Chem. B* **116**, 13024 (2012).

[29] Canongia Lopes, J. N. A.; Pádua, A. A. H. *J. Phys. Chem. B*, **110**, 7485–7489 (2006).

[30] E. Bodo, S. Mangialardo, P. Postorino, A. Sferrazza, and R. Caminiti, *, J. Chem. Phys.* **139**, 144309 (2013)

[31] E. Bodo, S. Mangialardo, F. Ramondo, F. Ceccacci and P. Postorino, *J. Phys. Chem. B* **116**, 13878 (2012).

[32] M. A. Ricci, M. Nardone, A. Fontana, C. Andreani and W. Hahn, *J. Chem. Phys.* **108**, 450 (1998).

[33] G. E. Walrafen, *J. Chem. Phys.* **40**, 3249 (1964).

[34] P. Postorino, A. Congeduti, E. Degiorgi, J. P. Itiè, and P. Munsch, *Phys. Rev. B* **65**, 224102 (2002).

[35] Pings, C. J.; Waser, J. *J. Chem. Phys.*, **1968,** *48*, 3016

[36] T. Pott and P. Meleard, *Phys. Chem. Chem. Phys.* **11**, 5469 (2009).

[37] C. Tanford, *J. Phys. Chem.* **76**, 3020 (1972).

[38] T. Head-Gordon and M. E. Johnson, *Proc. Natl. Acad. Sci. U.S.A* **103**, 7973 (2006).

[39] G. Hura, J. M. Sorenson, R. M. Glaeser and T. Head-Gordon, *J. Chem. Phys.* **113**, 9140 (2000).